\documentclass[12pt,aps,preprintnumbers,nofootinbib,showpacs]{revtex4}
\usepackage{amsmath,amssymb,bm,epsfig,slashbox}

\newcommand\eps{\epsilon}
\renewcommand\d{\partial}
\newcommand\grad{\bm{\nabla}}
\newcommand\p{{\bm{p}}}
\newcommand\q{{\bm{q}}}
\renewcommand\k{{\bm{k}}}
\newcommand\x{{\bm{x}}}
\newcommand\y{{\bm{y}}}
\newcommand\z{{\bm{z}}}
\newcommand\+{\dagger}
\renewcommand\O{{\mathcal{O}}}
\renewcommand\j{\bm{j}}
\newcommand\<{\langle}
\renewcommand\>{\rangle}
\newcommand\psiup{{\psi_\uparrow}}
\newcommand\psidown{{\psi_\downarrow}}
\renewcommand\a{\bm{a}}

\def\Jvol<#1,#2,#3>{#1}
\def\Jpage<#1,#2,#3>{#2}
\def\Jyear<#1,#2,#3>{#3}
\newcommand\journal[1]{\textbf{\Jvol<#1>}, \Jpage<#1> (\Jyear<#1>)}
\newcommand\PRA[1]{Phys.\ Rev.\ A \journal{#1}}
\newcommand\PRB[1]{Phys.\ Rev.\ B \journal{#1}}
\newcommand\PRD[1]{Phys.\ Rev.\ D \journal{#1}}
\newcommand\PLB[1]{Phys.\ Lett.\ B \journal{#1}}
\newcommand\ibid[1]{\textit{ibid.}\ \journal{#1}}

\begin{document}
\preprint{INT-PUB 07-16}

\title{Nonrelativistic conformal field theories}
\author{Yusuke~Nishida}
\affiliation{Institute for Nuclear Theory, University of Washington,
Seattle, Washington 98195-1550, USA}
\author{Dam~T.~Son}
\affiliation{Institute for Nuclear Theory, University of Washington,
Seattle, Washington 98195-1550, USA}

\begin{abstract}

We study representations of the Schr\"odinger algebra in terms of
operators in nonrelativistic conformal field theories.  We prove a
correspondence between primary operators and eigenstates of few-body
systems in a harmonic potential.  Using the correspondence we compute
analytically the energy of fermions at unitarity in a harmonic potential
near two and four spatial dimensions.  We also compute the energy of
anyons in a harmonic potential near the bosonic and fermionic limits.

\end{abstract}

\date{June 2007}


\maketitle

\section{Introduction}

Conformal field theories (CFTs) form a special class of relativistic
quantum field theories, where the Poincar\'e symmetry group is enlarged
to the group of conformal transformations.  One element of the
conformal algebra is dilatation: CFTs are always scale invariant.  The
conformal algebra and its representation have been studied
extensively~\cite{Fradkin:1997df}.

In this paper we study nonrelativistic counterparts of relativistic
conformal field theories.  There are several examples of such 
theories beside the trivial noninteracting theories.  Nonrelativistic
particles interacting through a $1/r^2$ potential is one such example.
The physically most important example in three spatial dimensions is
the theory describing spin-$1/2$ fermions with point-like interaction
fine-tuned to infinite scattering length (fermions at
unitarity)~\cite{Mehen:1999nd}.  Such fermionic systems have been
created and studied experimentally.  Theories describing anyons present
another class of nonrelativistic CFTs, but in two spatial dimensions.

The nonrelativistic analog of the conformal algebra is the so-called
Schr\"odinger algebra~\cite{Hagen:1972pd,Niederer:1972}.  While the
Schr\"odinger algebra has been studied
before~\cite{Hussin:1986cc,Jackiw:1990mb,Henkel:1993sg}, we are mostly
interested in the representations of the Schr\"odinger algebra in terms
of operators.  We show that the concept of primary operators can be
directly transferred to nonrelativistic theories.  In addition, we show
that there is an operator-state correspondence: a primary operator
(with some exceptions) corresponds to an eigenstate of a few-particle
system in a harmonic potential.  The scaling dimension of the operator
coincides with the energy of the corresponding eigenstate, divided by
the oscillator frequency.

The operator-state correspondence allows us to translate the problem of
finding the energy eigenvalues of a few-fermion state at unitarity, or a
few-anyon state, in a harmonic potential to another problem of finding
the anomalous dimensions of composite operators in the nonrelativistic
conformal field theory in free space.  The latter problem is amenable to
standard diagrammatic techniques for fermions at unitarity near two or
four spatial dimensions, or for anyons near the bosonic and fermionic
limits.  We present a few examples of such calculations in this paper.
In particular, we compute the ground state energy of up to six fermions
at unitarity in a harmonic potential near two and four dimensions, and
interpolate the results to find the energy in three dimensions.  We also
compute the ground state energy of up to four anyons in a harmonic
potential.

\section{Schr\"odinger algebra}

\subsection{Derivation of the algebra}

We briefly review the Schr\"odinger algebra~\cite{Hagen:1972pd,Niederer:1972}.
For definiteness, consider a nonrelativistic theory described by a
second-quantized field $\psi_\alpha(\x)$ (where $\alpha$ is the spin
index) which satisfies the commutation or anticommutation relation
\begin{equation}
  [\psi_\alpha(\x),\, \psi_\beta^\+(\y)]_\pm = \delta(\x-\y)
  \delta_{\alpha\beta}.
\end{equation}
Throughout this paper, we use nonrelativistic natural units $\hbar=m=1$
where $m$ is a particle mass.  We consider a general spatial dimension
$d$.  Define the number density and momentum density,
\begin{equation}
  n(\x) = \psi^\+(\x) \psi(\x), \qquad
  j_i(\x) = - \frac i2 (\psi^\+(\x) \d_i \psi(\x) - \d_i \psi^\+(\x) \psi(\x))
\end{equation}
(summation over spin indices is implied).  Their commutators are
\begin{subequations}\label{nj-comm}
\begin{align}
  & [n(\x),\, n(\y)] = 0, \qquad
    [n(\x),\, j_i(\y)] = -i n(\y) \d_i \delta(\x-\y),\\
  & [j_i(\x),\, j_j(\y)] = -i \left( j_j(\x)\d_i + j_i(\y)\d_j \right)
     \delta(\x-\y).
\end{align}
\end{subequations}

The Schr\"odinger algebra is formed by the following operators:
\begin{align}
  & N = \int\!d\x\, n(\x), \qquad
  P_i = \int\!d\x\, j_i(\x), \qquad
  M_{ij} = \int\! d\x\, (x_i j_j(\x) - x_j j_i(\x)),\label{NPM}\\
  & K_i = \int\!d\x\, x_i n(\x), \qquad
  C = \int\!d\x\, \frac{x^2}2 n(\x), \qquad
  D = \int\!d\x\, x_i j_i(\x),
\end{align}
and the Hamiltonian $H$.  The operators in Eq.~(\ref{NPM}) have simple
physical interpretation: $N$ is the particle number, $P_i$ is the
momentum, and $M_{ij}$ is the orbital angular momentum.  In a
scale-invariant theory like unitary fermions, these operators form a
closed algebra.  All commutators except those that involve $H$ can be
found from Eqs.~(\ref{nj-comm}). First $N$ commutes with all other
operators:
\begin{equation}
  [N,\, \textrm{any}] = 0.
\end{equation}
The commutator of the angular momentum $M_{ij}$ with an operator is
determined by the transformation properties of the latter under
rotations,
\begin{align}
  & [M_{ij},\, N] = [M_{ij},\, C] = [M_{ij},\, D] = 0,\\
  & [M_{ij},\, P_k] = i(\delta_{ik} P_j - \delta_{jk} P_i),\qquad
  [M_{ij},\, K_k] = i(\delta_{ik} K_j - \delta_{jk} K_i),\\
  & [M_{ij},\, M_{kl}]= i(\delta_{ik} M_{jk} - \delta_{jk} M_{il}
    + \delta_{il} M_{kj} - \delta_{jl} M_{ki}).
\end{align}
The remaining commutators are
\begin{align}
  & [P_i,\, P_j] = [K_i,\, K_j] = [K_i,\, C] = 0, \qquad
    [K_i,\, P_j] = i \delta_{ij}N,\\
  & [D,\, P_i] = iP_i, \qquad [D,\, K_i] = -i K_i, \qquad
    [D,\, C] = -2iC.
\end{align}
Now let us find the commutators of the Hamiltonian $H$ with other
operators.  Conservation of particle number, momentum, and angular
momentum implies that
\begin{equation}
  [H,\, N] = [H,\, P_i] = [H,\, M_{ij}] = 0.
\end{equation}
On the other hand, the continuity equation implies that
\begin{equation}
  [H,\, n] = -i\d_t n = i \d_i j_i,
\end{equation}
from which it follows that
\begin{equation}
  [H,\, K_i] = -i P_i, \qquad [H,\, C] = -i D.
\end{equation}
The computation of the last commutator $[H,\,D]$ requires the
condition of scale invariance.  For definiteness, let us consider fermions
at unitarity, described by the Hamiltonian
\begin{equation}
  H = \int\!d\x\, \frac12\d_i\psi^\+_\alpha\d_i\psi_\alpha
      + \frac12 \int\!d\x\,d\y\, \psi^\+_\alpha(\x) \psi^\+_\beta(\y)
      V(|\x-\y|) \psi_\beta(\y) \psi_\alpha(\x),
\end{equation}
where $V(|\x-\y|)$ is a short-range potential with infinite scattering
length.  We note that $D$ is an operator of dilatation,
\begin{equation}
  e^{-i\lambda D}\psi(\x)e^{i\lambda D} = e^{d\lambda/2}
  \psi(e^\lambda \x),
\end{equation}
from which one finds
\begin{equation}
  e^{-i\lambda D} H e^{i\lambda D} = e^{2\lambda} H',
\end{equation}
where $H'$ is the same as $H$ but the potential $V$ is replaced with a
new potential:
\begin{equation}
  V(r)\to V'(r) = e^{-2\lambda} V(e^{-\lambda}r).
\end{equation}
If $V$ corresponds to infinite scattering length, then $V'$ also
corresponds to infinite scattering length.  From the point of view of
long-distance physics, $H'=H$.  Therefore, we find
\begin{equation}\label{DH-comm}
 [D,\, H] = 2iH.
\end{equation}
It is clear that Eq.~(\ref{DH-comm}) is simply the condition of scale
invariance, and hence must hold for noninteracting anyons in two
spatial dimensions.

A more lengthy proof, which can be given for particles at unitarity,
is to use the momentum conservation equation
\begin{equation}
  \d_t j_i + \d_j \Pi_{ij} = 0,
\end{equation}
where $\Pi_{ij}$ is the stress tensor, which can be defined for a
generic potential $V(|\x-\y|)$ (see Appendix~\ref{sec:stress}).  The
commutator is then
\begin{equation}
  [D,\, H] = i \int\!d\x\, \Pi_{ii}(\x).
\end{equation}
At unitarity, one can show that (see Appendix~\ref{sec:stress})
\begin{equation}
  \int\!d\x\, \Pi_{ii}(\x) = 2 H,
\end{equation}
and Eq.~(\ref{DH-comm}) follows.

The Schr\"odinger algebra is summarized in Appendix~\ref{sec:algebra}.

\subsection{Local operators and representations of the Schr\"odinger algebra}

We introduce the notation of local operators $\O(t,\x)$ as
operators which depend on the position in time and space $t,\x$ so that
\begin{equation}
  \O(t,\x) = e^{iHt-iP_i x_i}\O(0) e^{-iHt+i P_i x_i}.
\end{equation}
A local operator $\O$ is said to have scaling dimension $\Delta_\O$ if
\begin{equation}
  [D,\, \O(0)] = i \Delta_\O \O(0),
\end{equation}
and to have particle number $N_\O$ if
\begin{equation}
  [N,\, \O(0)] = N_\O \O(0).
\end{equation}
We shall consider only operators with well-defined particle number and
scaling dimension.  Examples of such operators are $\psi$ and $\psi^\+$;
$\Delta_\psi=\Delta_{\psi^\+}=d/2$ and $-N_\psi=N_{\psi^\+}=1$.  In the
case of fermions at unitarity, a more complicated local operator is
\begin{equation}\label{phipsipsi}
  \phi(\x) = \lim_{\y\to\x} |\x-\y|^{d-2} \psi_\downarrow(\x) 
    \psi_\uparrow(\y).
\end{equation}
The presence of the prefactor $|\x-\y|^{d-2}$ guarantees that the
matrix elements of the operator $\phi(\x)$ between two states in the
Hilbert space are finite.\footnote{The condition of unitarity requires that
the wave function of $N$ spin-up and $M$ spin-down fermions 
$\Psi(\x_1,\x_2,\ldots,\x_N;\y_1,\y_2,\ldots,\y_M)$ behaves like
$|\x_i-\y_j|^{2-d}$ when $|\x_i-\y_j|\to0$ for any pair of fermions with
opposite spins $i,j$.}
The scaling dimension of $\phi$ is $\Delta_\phi=2$.
This can be found applying elementary dimension counting to
Eq.~(\ref{phipsipsi}):
\begin{equation}\label{dimphi}
  \Delta_\phi = 2\Delta_\psi + (d-2)\Delta_\x = 2\frac d2 + (d-2)(-1) = 2.
\end{equation}

Let us look at the set of all local operators $\O_a(t,\x)$.  These
operators, put at $t=0$ and $\x=0$, form a representation of the Schr\"odinger
algebra: for any operator $A$ in the algebra
\begin{equation}
  [A,\, \O_a(0)] = A_{ab} \O_b(0).
\end{equation}
We shall discuss the irreducible representations of the Schr\"odinger
algebra.

First we notice that if $\O$ has dimension $\Delta_\O$, then
$[P_i,\,\O]$ has dimension $\Delta_\O+1$:
\begin{equation}
 \begin{split}
  [D,\, [P_i,\, \O]] &= [P_i,\, [D,\, \O]] + [[D,\, P_i],\, \O] \\
  &= [P_i,\, i\Delta_\O\O] + [iP_i,\, \O] = i(\Delta_\O+1) [P_i,\,\O].
 \end{split}
\end{equation}
Analogously the dimensions of $[K_i\, \O]$, $[C,\, \O]$, and
$[H,\,\O]$ are $\Delta_\O-1$, $\Delta_\O-2$, and $\Delta_\O+2$,
respectively:
\begin{align}
  [D,\, [K_i,\, \O]] &= i(\Delta_\O-1)[K_i,\, \O],\\
  [D,\, [C,\, \O]] &= i(\Delta_\O-2)[C,\, \O],\\
  [D,\, [H,\, \O]] &= i(\Delta_\O+2)[H,\, \O].
\end{align}
Assuming that the dimensions of operators are bounded from below, if
one starts with a given operator and repeatedly takes its commutator
with $K_i$ and $C$, one lowers the dimension of the operator until it
cannot be lowered further.  The last operator $\O$ obtained this way
has the property
\begin{equation}
  [K_i,\, \O] = [C,\, \O] = 0.
\end{equation}
Operators that commute with $K_i$ and $C$ will be called \emph{primary
operators} (quasiprimary operators in the terminology of
Ref.~\cite{Henkel:1993sg}).  In general, $[K_i,\, \O]=0$ does not
imply that $[C,\, \O]=0$, and vice versa.

Starting with a primary operator $\O$, one can build up a tower of
operators by taking commutators with $P_i$ and $H$.  In other words,
starting with primary operators one can build up whole towers of
operators by taking their space and time derivatives.  For example, the
operators with dimension $\Delta_\O+1$ in the tower are 
$[P_i,\, \O]\equiv i\d_i\O$.  At the next level (dimension
$\Delta_\O+2$), the following are possible:
\begin{equation}\label{dtdidjO}
  [H,\, \O] = -i\d_t\O, \quad [P_i,\, [P_j,\, \O]] \equiv - \d_i\d_j\O.
\end{equation}
Commuting the operators in Eq.~(\ref{dtdidjO}) with $K_i$ and $C$, we
get back the operators in the lower rungs of the tower.

It is easy to see that the operators built from a primary operator by
taking space and time derivatives form an irreducible representation
of the Schr\"odinger algebra.  It is also possible to show that the
full set of all local operators can be decomposed into irreducible
representations, each of which is built upon a single primary
operator.  The task of finding the spectrum of dimensions of all local
operators reduces to finding the dimensions of primary operators.

For an operator $\O(t,\x)$ with dimension $\Delta_\O$ at an arbitrary
spacetime point, the following commutation relations hold,
\begin{align}
  & [P_i,\, \O] = i\d_i \O, \qquad  [H,\, \O] = -i \d_t \O,\\
  & [D,\, \O] = i (2t\d_t + x_i\d_i + \Delta_\O) \O. \label{DOx}
\end{align}
Moreover, if $\O$ is a primary operator then
\begin{align}\label{KCO}
  [K_i,\, \O] &= (-it\d_i + N_\O x_i) \O,\\
  [C,\, \O] &= -i(t^2\d_t + tx_i\d_i + t\Delta_\O) \O + \frac{x^2}2N_\O\O.
\end{align}
The exponentiated version of Eq.~(\ref{DOx}) is
\begin{equation}\label{DOx-exp}
  e^{-i\lambda D} \O(t,\x) e^{i\lambda D} = 
  e^{\lambda\Delta_\O} \O(e^{2\lambda}t, e^\lambda \x).
\end{equation}

For any set of $n$ operators, one can define an $n$-point correlation
function,
\begin{equation}
  G_n(t_1,\x_1, t_2,\x_2,\ldots, t_n,\x_n) = 
  \<0 | T \O_1(t_1,\x_1) \O_2(t_2,\x_2) \ldots \O_n(t_n,\x_n) |0\> ,
\end{equation}
where $T$ is time ordering.  Clearly for $G_n$ to be nonzero it is
necessary that $N_{\O_1}+N_{\O_2}+\cdots+N_{\O_n}=0$.  If all $\O_i$
have definite dimensions then this correlation function has a scaling
property
\begin{equation}
  G_n(e^{2\lambda}t_i, e^\lambda \x_i) = 
  \exp\left(-\lambda \sum_{i=1}^n \Delta_{\O_i}\right)
  G_n(t_i, \x_i),
\end{equation}
which follows from Eq.~(\ref{DOx-exp}) and $e^{i\lambda D}|0\>=|0\>$.

The correlation functions of primary operators are further
constrained~\cite{Henkel:1993sg}.  As an example, consider the two-point
correlation function of a primary operator $\O$ with its Hermitian
conjugate:
\begin{equation}
  G(t,\x) = \<0| T \O(t,\x) \O^\+(0)|0\> .
\end{equation}
Using $\<0|[K_i,\, T \O(x) \O^\+(y)]|0\>=0$ and Eqs.~(\ref{KCO}) one
obtains
\begin{equation}
  (-it\d_i + N_\O x_i) G(t,\x) = 0 .
\end{equation}
Combining with the scale invariance, the two-point correlation function
is determined up to an overall coefficient,
\begin{equation}
  G(t,\x) = C t^{-\Delta_\O} \exp\left( -iN_\O \frac{|\x|^2}{2t}\right).
\end{equation}

\subsection{Correspondence to states in a harmonic potential}

We now show that each primary operator corresponds to an energy
eigenstate of a system in a harmonic potential.  We set the oscillator
frequency of the harmonic potential $\omega$ to $1$.  The total
Hamiltonian of the system in a harmonic potential is
\begin{equation}
  H_{\textrm{osc}} = H + C.
\end{equation}

Consider a primary operator $\O$ put at $t=0$ and $\x=0$.  Let $\O$ be
constructed from annihilation operators, so that $\O^\+$ acts
nontrivially on the vacuum $|0\>$.  Consider the following state
\begin{equation}
  |\Psi_\O\> = e^{-H} \O^\+ |0\>.
\end{equation}
If the particle number of $\O^\+$ is $N_{\O^\+}$, then $|\Psi_\O\>$ is an
$N_{\O^\+}$-body state.  Let us show that $|\Psi_\O\>$ is an eigenstate
of the Hamiltonian: $H_{\textrm{osc}} = H + C$.  Indeed
\begin{equation}
  H_{\textrm{osc}}|\Psi_\O\> = e^{-H} (e^H H_{\textrm{osc}} e^{-H})
    \O^\+ |0\>.
\end{equation}
We now use the formula
\begin{equation}
  e^H H_{\textrm{osc}} e^{-H} = H_{\textrm{osc}} 
  + [H,\, H_{\textrm{osc}}] + \frac12 [H,\, [H,\, H_{\textrm{osc}}]]
  + \cdots .
\end{equation}
Using the commutation relations in Appendix~\ref{sec:algebra}, we find
that all terms in the $\cdots$ vanish, and the right hand side is equal
to $C-iD$.  Therefore
\begin{equation}
  H_{\textrm{osc}}|\Psi_\O\> = e^{-H}(C-iD)\O^\+ |0\> = 
  e^{-H} \O^\+ (C-iD)|0\> + e^{-H} [C-iD,\, \O^\+] |0\>.
\end{equation}
However, both $C$ and $D$ annihilate the vacuum, $C|0\>=D|0\>=0$, and
since $\O$ is a primary operator, $[C,\, \O^\+]=0$.  Thus, using
$[D,\, \O^\+] = -[D,\, \O]^\+ = i\Delta_\O \O^\+$, we obtain
\begin{equation}
  H_{\textrm{osc}}|\Psi_\O\> = e^{-H}\Delta_\O \O^\+ |0\> 
   = \Delta_\O |\Psi_\O\>,
\end{equation}
i.e., $|\Psi_\O\>$ is an eigenstate of the system of $N_{\O^\+}$
particles in a harmonic potential, with the energy eigenvalue
$\Delta_\O$ (times $\hbar\omega$).

It is known that the eigenstates of $H_{\textrm{osc}}$ are organized into
ladders with spacing between steps equal to 2~\cite{WernerCastin,Tan}.
The raising and lowering operators within a ladder
are~\cite{WernerCastin}
\begin{align}
  L_+ &= H - C + iD,\\
  L_- &= H - C - iD.
\end{align}
Let us show that the state $|\Psi_\O\>$ is annihilated by $L_-$ and
hence is the lowest state in its ladder.  Indeed, using the identity
\begin{equation}
  e^H L_- e^{-H} = -C,
\end{equation}
we find
\begin{equation}
  L_- |\Psi_\O\> = e^{-H}\, e^H L_- e^{-H}\, \O^\+|0\> =
  - e^{-H} C \O^\+ |0\> = -e ^{-H} \O^\+ C|0\> = 0.
\end{equation}

Clearly, in order to correspond to a nontrivial eigenstate of 
$H_{\textrm{osc}}$, $\O^\+$ must not annihilate the vacuum:
$\O^\+|0\>\neq0$.  We shall consider the operators $\O$ that are built
from the fundamental annihilation operators of the field theories.

\subsection{Simple examples: one and two-body operators/states}

Let us illustrate this correspondence using one-particle and two-particle
operators $\psi$ and $\phi$ at unitarity.  The operator $\psi$ has
scaling dimension $d/2$, which matches the ground state energy of one
particle in a harmonic potential in spatial dimension $d$.  The operator
$\phi$ has scaling dimension $2$.  The ground state of two particles at
unitarity in a harmonic potential has the wave function
\begin{equation}\label{2bodywf}
  \phi(\x,\y) \propto \frac{e^{-(x^2+y^2)/2}}{|\x-\y|^{d-2}},
\end{equation}
and the ground state energy is also 2.

\section{Example 1: Fermions at unitarity}

In this section, we compute the scaling dimensions of some operators in
the theory describing spin-$1/2$ fermions at unitarity.  In order to
have a small parameter for perturbative expansions, we shall work near
two and four spatial dimensions, and then, interpolate the results to
the physical three spatial dimensions.  Since the energy eigenvalues of
two and three fermions in a harmonic potential can be found exactly, we
can use these cases to test our expansions and interpolation schemes.
In the cases of more than three fermions, only numerical results exist.
Our analytical calculations, as we will see, are consistent with the
numerical ones.

There are two field-theoretical representations of fermions at
unitarity, one becoming weakly coupled as $d\to4$ and the other becoming
weakly coupled as $d\to 2$~\cite{Nishida-Son06,Nikolic-Sachdev07}.  We
shall consider these two cases separately.

\subsection{Near four spatial dimensions}

\subsubsection{Fixed point}

In the first representation, the Lagrangian density describing fermions
at unitarity is
\begin{equation}\label{L-4d}
 \mathcal{L} 
  = i\psi_\sigma^\+\d_t \psi_\sigma - \frac12|\nabla\psi_\sigma|^2
  + i\phi^*\d_t\phi - \frac14 |\nabla\phi|^2 
  + g\psi_\uparrow^\+ \psi_\downarrow^\+ \phi + g\psidown \psiup \phi^*.
\end{equation}
The canonical dimensions of the fermion field $\psi_\sigma$ and the
boson field $\phi$ are both $d/2$.  Therefore the coupling constant $g$
is relevant at weak coupling below $d=4$.  There are three other
relevant terms one can add to the Lagrangian density~(\ref{L-4d}):
$\mu_\sigma\psi_\sigma^\+\psi_\sigma$ and 
$-\left(g^2/c_0\right)\phi^*\phi$.  $\mu_\sigma$ is a chemical potential
for each spin component of fermions and here we consider the system at zero
density $\mu_\sigma=0$.  Furthermore we assume that the system is
fine-tuned so that the coefficient in front of $\phi^*\phi$ satisfies
\begin{equation}
 \frac1{c_0} = \int\!\frac{d\k}{(2\pi)^d}\frac1{k^2}.
\end{equation}
[$(c_0)^{-1}$ is zero in dimensional regularization.]  This
condition is equivalent to the fine-tuning to the infinite scattering
length.  We denote the propagators of $\psi$ and $\phi$ by $G(p)$ and
$D(p)$, respectively.

\begin{figure}[tp]
 \includegraphics[width=0.55\textwidth,clip]{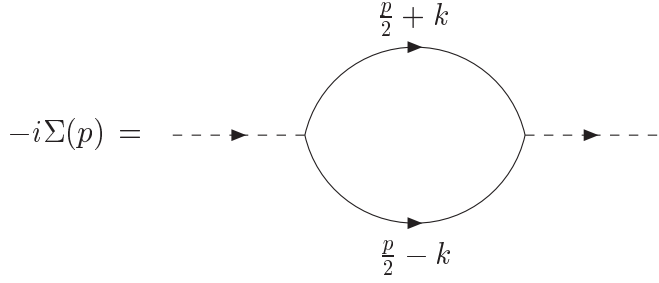}
 \caption{One-loop self-energy diagram to renormalize the wave function
 of $\phi$.  \label{fig:self-energy}}
\end{figure}

The renormalization of the theory can be performed in the standard
way considering $\eps=4-d$ to be a small parameter for perturbation.
There is a one-loop self-energy diagram for $\phi$ which is
logarithmically divergent (Fig.~\ref{fig:self-energy}).  Integrating out
modes in the momentum shell $e^{-s}\Lambda<k<\Lambda$, we obtain 
\begin{equation}
 \begin{split}
  \Sigma(p) &= \frac{g^2}{c_0} + i g^2 \int\!\frac{dk}{(2\pi)^{d+1}}
  G\!\left(\frac p2+k\right)G\!\left(\frac p2-k\right) \\
  &= -\frac{g^2}{8\pi^2}\left( p_0 - \frac{p^2}4\right) 
  \ln\frac\Lambda{e^{-s}\Lambda},
 \end{split}
\end{equation}
which corresponds to the wave-function renormalization of $\phi$
\begin{equation}\label{Zphi}
  Z_\phi = 1 - \frac{g^2}{8\pi^2} s.
\end{equation}
The anomalous dimension of $\phi$ is found by the standard formula
\begin{equation}
 \gamma_\phi = -\frac12 \frac{\d\ln Z_\phi}{\d s} = \frac{g^2}{16\pi^2}.
\end{equation}

There is no divergent one-particle irreducible diagram that renormalizes
the $\psidown\psiup\phi^*$ coupling.  As a result, the $\beta$ function
that governs the running of $g$, 
\begin{equation}
  \frac{\d g}{\d s} = \beta(g), 
\end{equation}
is determined by the dimension of $g$:
\begin{equation}
  \beta(g) = \left( 2 - \frac d2 - \gamma_\phi \right) g
   = \frac\eps2 g - \frac{g^3}{16\pi^2}.
\end{equation}
There is a fixed point located at
\begin{equation}
  g^2 = 8\pi^2\eps.
\end{equation}
At this fixed point, the theory is a nonrelativistic CFT describing
fermions at unitarity.

\subsubsection{Scaling dimensions of operators}

Since the one-fermion operator $\psi$ is not renormalized, its scaling
dimension is
\begin{equation}
  \Delta_\psi = \frac d2.
\end{equation}
Using the operator-state correspondence, we find that there is a
one-fermion state in a harmonic potential with energy
$\left(d/2\right)\omega$, which is obvious.

The two-fermion operator $\phi$, on the other hand, has a scaling
dimension different from its canonical dimension $d/2$.  At the fixed
point,
\begin{equation}\label{Deltaphi}
  \Delta_\phi = \frac d2 + \gamma_\phi = 2.
\end{equation}
This is, of course, consistent with Eq.~(\ref{dimphi}).  Since there is
no other contribution to $\Delta_\phi$, Eq.~(\ref{Deltaphi}) is exact to
all order in $\eps$.  According to the operator-state correspondence, 
this implies the existence of a two-fermion state with zero orbital
angular momentum and energy $2\,\omega$.  The wave function of this
state is given by Eq.~(\ref{2bodywf}). 

\begin{figure}[tp]
 \includegraphics[width=0.5\textwidth,clip]{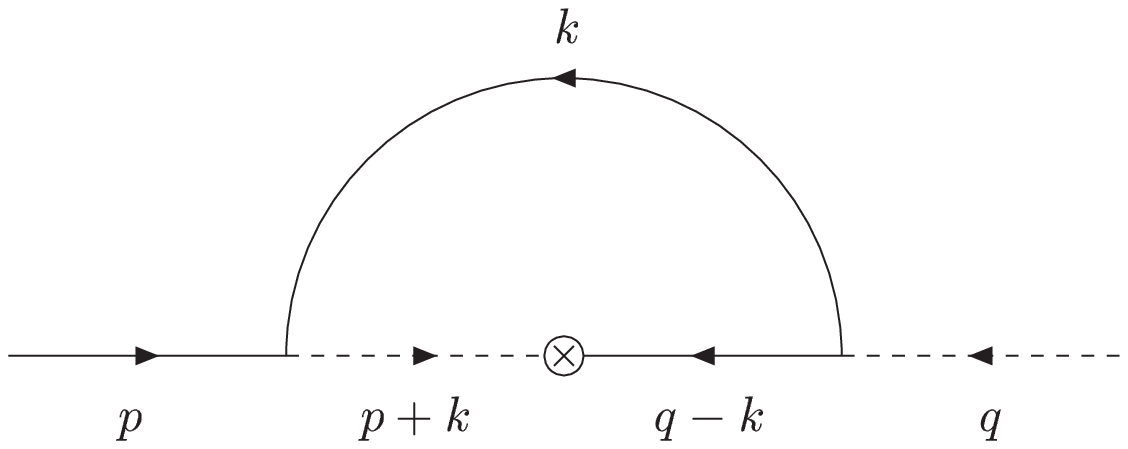}
 \caption{One-loop diagram to renormalize the three-fermion operator
 $\phi\psiup$.  \label{fig:boson-fermion}}
\end{figure}

Let us look at three-fermion operators.  The simplest operator is
$\phi\psiup$.  This operator has zero orbital angular momentum $l=0$.
The diagram that contributes to its anomalous dimension to order $\eps$
is depicted in Fig.~\ref{fig:boson-fermion}.  It is evaluated as
\begin{equation}
 -i g^2\int\!\frac{dk}{(2\pi)^{d+1}} G(k)D(k)G(-k)
 = -\frac{g^2}{6\pi^2} \ln\frac\Lambda{e^{-s}\Lambda}.
\end{equation}
Therefore the renormalized operator differs from the bare operator by
a rescaling factor: 
$(\phi\psiup)_\mathrm{ren}=Z_{\phi\psiup}^{-1}\phi\psiup$, where 
\begin{equation}
  Z_{\phi\psiup} = 1 - \frac{g^2}{6\pi^2} s.
\end{equation}
At the fixed point, $-\d\ln Z_{\phi\psiup}/\d s=4\eps/3$ is the
anomalous dimension of the composite operator $\phi\psiup$ (more
precisely, the nontrivial part of the anomalous dimension because there
is a trivial part equal to $\gamma_\phi$).  We thus find the scaling
dimension
\begin{equation}
  \Delta_{\phi\psiup} = \Delta_\phi + \Delta_\psiup + \frac43 \eps
  = 4 + \frac56\eps.
\end{equation}
According to the operator-state correspondence, $\phi\psiup$ corresponds
to a three-fermion state with $l=0$ and energy equal to
\begin{equation}
 E_3^{(0)}=\left[4+\frac56\eps+O(\eps^2)\right]\omega.
\end{equation}
This state continues to the first excited state of three fermions in a
harmonic potential at $d=3$.  Even within the leading correction in
$\eps$, the result extrapolated to $\eps=1$,
$E_3^{(0)}\approx4.83\,\omega$, is not too far from the true result
of $4.66622\,\omega$ at $d=3$.

The three-fermion ground state in a harmonic potential at $d=3$ has
$l=1$.  There are two lowest $l=1$ operators near four spatial
dimensions; $\phi\grad\psiup$ and $(\grad\phi)\psiup$.  Let us consider
the renormalization of a general operator
$a\,\phi\grad\psiup+b\,(\grad\phi)$.  Inserting this operator into
Fig.~\ref{fig:boson-fermion}, we find
\begin{equation}
 \begin{split}
  &-i g^2 \int\!\frac{dk}{(2\pi)^{d+1}}
  \left[a\left(-\k+\q\right) + b\left(\k+\p\right)\right]
  G(k)D(k+p)G(-k+q) \\
  &= -\frac{g^2}{6\pi^2} \left[\frac{a+5b}{6}\p
  + \frac{5a+7b}{12}\q\right] \ln\frac\Lambda{e^{-s}\Lambda}.
 \end{split}
\end{equation}
In order to have well-defined anomalous dimensions, this should be
proportional to $a\p+b\q$.  Thus we have two solutions
$(a,b)\propto(1,1)$ and $(a,b)\propto(2,-1)$, which have anomalous
dimensions $4\eps/3$ and $-\eps/3$, respectively.  The first possibility
corresponds to the operator $\grad(\phi\psiup)$, that is obviously not a
primary operator.  Its scaling dimension is trivially equal to
$\Delta_{\phi\psiup}+1$ (corresponding to an excitation in the center of
mass motion).  The other operator $2\phi\grad\psiup-(\grad\phi)\psiup$
has the nontrivial scaling dimension
\begin{equation}
 \Delta_{2\phi\grad\psiup-(\grad\phi)\psiup}
  =\Delta_\phi+\Delta_\psiup+1-\frac13\eps=5-\frac56\eps.
\end{equation}
The operator-state correspondence tells us that the three-fermion state
with $l=1$ in a harmonic potential has the energy
\begin{equation}
 E_3^{(1)}=\left[5-\frac56\eps+O(\eps^2)\right]\omega.
\end{equation}
The extrapolation to $\eps=1$ gives $E_3^{(1)}\approx4.17\,\omega$,
which is not too far from the true ground state energy $4.27272\,\omega$
at $d=3$.

\begin{figure}[tp]
 \includegraphics[width=0.55\textwidth,clip]{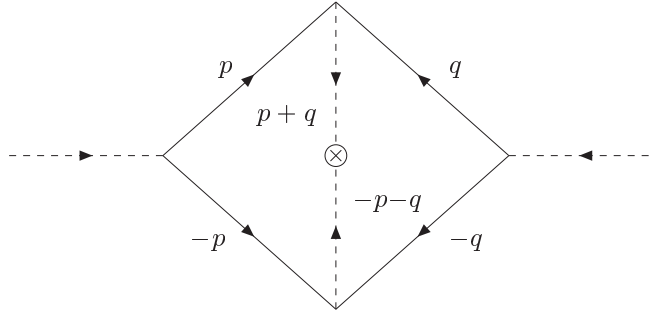}
 \caption{One-loop diagram to renormalize the four-fermion operator
 $\phi^2$.  \label{fig:boson-boson}}
\end{figure}

We now turn to the four-fermion state with $l=0$ represented by the
operator $\phi^2$.  The first nontrivial correction to its scaling
dimension is of order $\eps^2$ given by the diagram depicted in
Fig.~\ref{fig:boson-boson}.  The two-loop integral can be performed
analytically and we find
\begin{equation}\label{Deltaphi2}
  \Delta_{\phi^2} = 4 + 8 \eps^2 \ln\frac{27}{16}.
\end{equation}
Thus the ground state of four fermions in a harmonic potential has the
energy 
\begin{equation}
 E_4^{(0)}=\left[4+8\eps^2\ln\frac{27}{16}+O(\eps^3)\right]\omega.
\end{equation}
The correction, although it is of order $\eps^2$, has the large
coefficient.  Therefore, in order to extrapolate to $\eps=1$, we shall
not use Eq.~(\ref{Deltaphi2}) directly but will combine it with a result
near two spatial dimensions.

For a general even number of fermions $N=2n$, the operator $\phi^n$
corresponds to the ground state in a harmonic potential with $l=0$.  Its
scaling dimension is given by
\begin{equation}
  \Delta_{\phi^n} = N + N\left(N-2\right) \eps^2 \ln\frac{27}{16}
   + O(\eps^3).
\end{equation}
When $N=2n+1$ is odd, the operators $\phi^n\psiup$ and
$2\phi^n\grad\psiup-\phi^{n-1}(\grad\phi)\psiup$ correspond to states
with orbital angular momentum $l=0$ and $l=1$, respectively.  Their
scaling dimensions are
\begin{equation}
 \Delta_{\phi^n\psiup} = N+1 + \frac{4N-7}6 \eps + O(\eps^2)
\end{equation}
and
\begin{equation}
 \Delta_{2\phi^n\grad\psiup-\phi^{n-1}(\grad\phi)\psiup}
  = N+2 + \frac{2N-21}{18} \eps + O(\eps^2).
\end{equation}
According to the operator-state correspondence, the energy of
$N$-fermion state in a harmonic potential is simply given by
$E_N^{(l)}=\Delta_\O\,\omega$.

The leading-order results [$E_N^{(0)}=N\,\omega$ for even $N$ and 
$E_N^{(0)}=\left(N+1\right)\omega$ and
$E_N^{(1)}=\left(N+2\right)\omega$ for odd $N$] can be easily
understood by recalling that, in the limit of $d\to4$ from below,
fermion pairs at unitarity form point-like bosons and they do not
interact with each other or with extra
fermions~\cite{Nishida-Son06,Nussinov06}.  So the ground state for $N=2n$
fermions consists of $n$ free composite bosons, each of which has the
lowest energy $2\,\omega$ in a harmonic potential at $d=4$.  When
$N=2n+1$, the ground state has $l=0$ and consists of $n$ composite
bosons and one extra fermion in the lowest energy states.  In order to
have an $l=1$ state, one of the $n+1$ particles has to be excited to the
first excited state, which costs additional $1\,\omega$.  At $d=4$,
we observe the odd-even staggering in the ground state energy as
$E_N^{(0)}-\bigl(E_{N-1}^{(0)}+E_{N+1}^{(0)}\bigr)/2=1\,\omega$
for odd $N$.

\subsection{Near two spatial dimensions}

\subsubsection{Fixed point}

\begin{figure}[tp]
 \includegraphics[width=0.45\textwidth,clip]{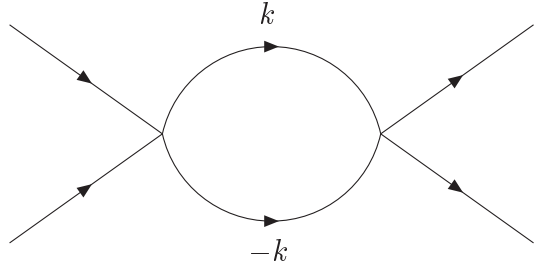}
 \caption{One-loop diagram to renormalize the four-point vertex
 coupling $\bar g^2$.  \label{fig:vertex}}
\end{figure}

The other field-theoretical representation of fermions at unitarity is
provided by the Lagrangian density
\begin{equation}
  \mathcal{L} = i\psi_\sigma^\+\d_t \psi_\sigma 
  - \frac12|\nabla\psi_\sigma|^2 + \bar g^2
  \psi_\uparrow^\+ \psi_\downarrow^\+ \psidown \psiup.
\end{equation}
Again we consider the system at zero density $\mu_\sigma=0$.  
When $\bar\eps=d-2$ is small, the coupling $\bar g^2$ is renormalized by
the logarithmically divergent one-loop diagram (Fig.~\ref{fig:vertex}).
The renormalization group equation for $\bar g^2$ is
\begin{equation}
  \frac{\d \bar g^2}{\d s}
   = - \bar\eps \bar g^2 + \frac{\bar g^4}{2\pi}.
\end{equation}
The fixed point is located at
\begin{equation}
  \bar g^2 = 2\pi\bar\eps.
\end{equation}
At this fixed point, the theory is a nonrelativistic CFT describing
fermions at unitarity.

\subsubsection{Scaling dimensions of operators}

\begin{figure}[tp]
 \includegraphics[width=0.35\textwidth,clip]{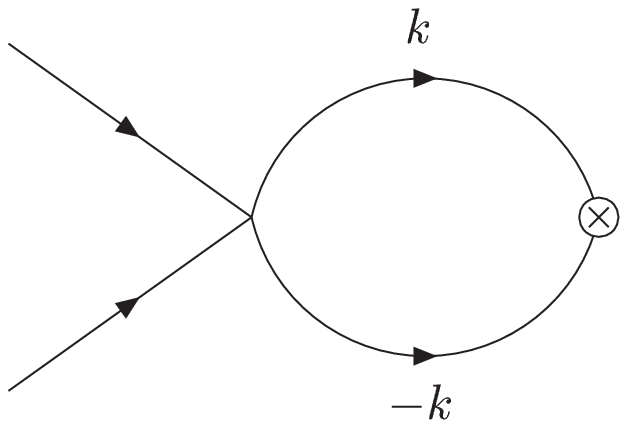}
 \caption{One-loop diagram to renormalize the two-fermion operator
 $\psidown\psiup$.  \label{fig:fermion-fermion}}
\end{figure}

The scaling dimensions of the one-fermion operator $\psi$ and the
two-fermion operator $\psidown\psiup$ (its renormalization is given by
Fig.~\ref{fig:fermion-fermion}) are given by the same formulas as around
four dimensions; $\Delta_\psi=d/2$ and $\Delta_{\psidown\psiup}=2$.
Here we concentrate our attention on three-fermion operators.  The
simplest operator is $\psiup\psidown\grad\psiup$, which has the orbital
angular momentum $l=1$.  By considering the diagram where one more
fermion line is attached to Fig.~\ref{fig:fermion-fermion}, the
renormalization of this operator is given by
\begin{equation}
\begin{split}
  &-i\bar g^2 \int\!\frac{dk}{(2\pi)^{d+1}} 
  \left(\p_3 - \frac{\p_1+\p_2}2-\k \right) 
  G\!\left(\frac{p_1+p_2}2+k\right) G\!\left(\frac{p_1+p_2}2-k\right) \\
  &-i\bar g^2 \int\!\frac{dk}{(2\pi)^{d+1}} 
  \left(\frac{\p_2+\p_3}2+\k - \p_1\right)
  G\!\left(\frac{p_2+p_3}2+k\right) G\!\left(\frac{p_2+p_3}2-k\right) \\
  &=\frac{3\bar g^2}{4\pi} \left(\p_2-\p_1\right)
 \ln\frac\Lambda{e^{-s}\Lambda}.
 \end{split}
\end{equation}
Therefore the renormalized operator is
$(\psiup\psidown\grad\psiup)_\mathrm{ren}=(Z_{\psiup\psidown\grad\psiup})^{-1}\psiup\psidown\grad\psiup$,
where
\begin{equation}
 Z_{\psiup\psidown\grad\psiup}=1+\frac{3\bar g^2}{4\pi}s.
\end{equation}
At the fixed point, the anomalous dimension becomes 
$\gamma_{\psiup\psidown\grad\psiup}=-\d\ln Z_{\psiup\psidown\grad\psiup}/\d s=-3\bar\eps/2$.  
So the scaling dimension of the operator $\psiup\psidown\grad\psiup$ is
\begin{equation}
  \Delta_{\psiup\psidown\grad\psiup} 
   = \frac{3d}2+1+\gamma_{\psiup\psidown\grad\psiup} = 4.
\end{equation}
According to the operator-state correspondence, the ground state energy
of three fermions in a harmonic potential is given by
\begin{equation}
 E_3^{(1)} = \left[4+O(\bar\eps^2)\right]\omega.
\end{equation}

For three-fermion operators with $l=0$, the calculation is somewhat more
involved, because there are three lowest operators that can mix with
each other: $\psiup\psidown\nabla^2\psiup$,
$\psiup\grad\psidown{\cdot}\grad\psiup$, and $\psiup\psidown\d_t\psiup$.
The linear combinations with well-defined anomalous dimensions are
\begin{itemize}
\item $\grad{\cdot}(\psiup\psidown\grad\psiup)$ with
      $\gamma=-3\bar\eps/2$,
\item $\psiup\psidown\d_t\psiup$ with $\gamma=-3\eps/2$,
\item $\psiup\psidown\left(i\d_t+\tfrac12{\nabla^2}\right)\psiup$ with
      $\gamma=-\bar\eps$. 
\end{itemize}
The first operator is not a primary operator.  Its scaling dimension is
trivially equal to $\Delta_{\psiup\psidown\grad\psiup}+1$ (corresponding
to an excitation of the center of mass motion).  The third operator
annihilates the vacuum and thus does not correspond to any eigenstate of
the system in a harmonic potential.  The second operator is, therefore,
the one that corresponds to the lowest energy eigenstate of three 
fermions in a harmonic potential with $l=0$.  The energy of this state
is
\begin{equation}
 E_3^{(0)} = \left[\frac{3d}2+2 - \frac32\bar\eps\right]\omega
  = \left[5 + O(\bar\eps^2)\right]\omega.
\end{equation}

We can develop the same analysis for operators having more than three
fermion numbers.  The lowest four-fermion operator with $l=0$ is
$\psiup\psidown\grad\psiup{\cdot}\grad\psidown$.  Its anomalous
dimension is computed to be $-3\bar\eps$, which corresponds to the
ground state energy in a harmonic potential
\begin{equation}
  E_4^{(0)} = \left[6-\bar\eps + O(\bar\eps^2)\right]\omega.
\end{equation}

For five fermions, the operator
$\psiup\psidown(\grad\psiup{\cdot}\grad\psidown)\grad\psiup$ with the
anomalous dimension $-7\bar\eps/2$ corresponds to the ground state in a
harmonic potential with $l=1$.  Its energy is given by
\begin{equation}
 E_5^{(1)} = \left[8 - \bar\eps + O(\bar\eps^2)\right]\omega.
\end{equation}
We can find two nontrivial operators with $l=0$ corresponding to the
lowest two energy eigenstates of five fermions in a harmonic potential.
The operators $a\,\psiup\psidown(\grad\psiup{\cdot}\grad\psidown)\nabla^2\psiup
+b\,\psiup\nabla_{\!i\,}\psidown(\grad\psiup{\cdot}\grad\psidown)\nabla_{\!i\,}\psiup
+c\,\psiup\psidown((\nabla_{\!i\,}\grad\psiup){\cdot}\grad\psidown)\nabla_{\!i\,}\psiup
-d\,\psiup\psidown(\grad\psiup{\cdot}\grad\psidown)i\d_t\psiup$
with
\begin{equation}\label{eq:N5_s-wave}
 (a,b,c,d)\propto\left(\pm19\sqrt{3}-5\sqrt{35},\,\mp16\sqrt{3},\,-6\sqrt{35}\mp6\sqrt{3},\,16\sqrt{35}\right)
\end{equation}
have well-defined anomalous dimensions
$-\left(51\pm\sqrt{105}\right)\bar\eps/16$.  Therefore, there are
five-fermion states with $l=0$ and energies equal to
\begin{equation}
 E_5^{(0)} = 
  \left[9-\frac{11\pm\sqrt{105}}{16}\bar\eps+O(\bar\eps^2)\right]\omega.
\end{equation}

Finally, the lowest six-fermion operator with $l=0$ is
$\psiup\psidown(\grad\psiup{\cdot}\grad\psidown)(\grad\psiup{\cdot}\grad\psidown)$.
Its anomalous dimension is computed to be $-5\bar\eps$, which corresponds to the
ground state energy in a harmonic potential
\begin{equation}
  E_6^{(0)} = \left[10-2\bar\eps + O(\bar\eps^2)\right]\omega.
\end{equation}

We note that the leading-order results for $E_N^{(l)}$ can be easily
understood by recalling that, in the limit of $d\to2$ from above,
fermions at unitarity become noninteracting~\cite{Nishida-Son06,Nussinov06}.
So the energy eigenvalue of each $N$-fermion state is just a sum of
single particle energies in a harmonic potential at $d=2$.  Clearly, the
ground state energy shows the shell structure at $d=2$.

\subsection{Interpolations to $d=3$ and discussion}

\begin{table}[tp]
 \caption{Scaling dimensions of operators corresponding to $N$-fermion
 states with orbital angular momentum $l$ in a harmonic potential.
 Known results for the energy eigenvalues $E_N^{(l)}$ at $d=3$ are also
 shown in units of $\hbar\omega$~\cite{Tan04,Chang07,Stecher07}.
 \label{tab:dimensions}}
 \begin{ruledtabular}
 \begin{tabular}{l|l|l|l}
  \phantom{fermion number}&$\bar\eps=d-2$ expansion&$\eps=4-d$ expansion
  &known value at $d=3$\\\hline\hline
  \ $N=2$\ \ ($l=0$)&$2$&$2$&$2$\\\hline
  \ $N=3$\ \ ($l=0$)&$5+O(\bar\eps^2)$
  &$4+\frac56\eps+O(\eps^2)$&$4.66622$\\\hline
  \ $N=3$\ \ ($l=1$)&$4+O(\bar\eps^2)$
  &$5-\frac56\eps+O(\eps^2)$&$4.27272$\\\hline
  \ $N=4$\ \ ($l=0$)&$6-\bar\eps+O(\bar\eps^2)$
  &$4+8\eps^2\ln\frac{27}{16}+O(\eps^3)$&$5.1\pm0.1$\ /\ $5.07\pm0.01$\\\hline
  \ $N=5$\ \ ($l=0$)&$9-\frac{11+\sqrt{105}}{16}\bar\eps+O(\bar\eps^2)$
  &$6+\frac{13}6\eps+O(\eps^2)$&---\\\hline
  \ $N=5$\ \ ($l=1$)&$8-\bar\eps+O(\bar\eps^2)$
  &$7-\frac{11}{18}\eps+O(\eps^2)$&$7.6\pm0.1$\\\hline
  \ $N=6$\ \ ($l=0$)&$10-2\bar\eps+O(\bar\eps^2)$
  &$6+24\eps^2\ln\frac{27}{16}+O(\eps^3)$&$8.7\pm0.1$\ /\ $8.67\pm0.03$
 \end{tabular}
 \end{ruledtabular}
\end{table}

We determined the exact scaling dimensions of the one-fermion operator
$\psi$ and the two-fermion operator $\phi=\psidown\psiup$ in arbitrary
spatial dimension $d$.  For the scaling dimensions of $N$-fermion
operators with $N\geq3$, a few lowest-order terms in the expansions over
$\bar\eps=d-2$ and $\eps=4-\eps$ were computed as summarized in
Table~\ref{tab:dimensions}.  According to the operator-state
correspondence, we find that the ground state of three fermions in a
harmonic potential has the orbital angular momentum $l=1$ near $d=2$,
while $l=0$ near $d=4$.  So there must be at least one level crossing
between the states with $l=0$ and $l=1$ as $d$ increases.   Using the
$\eps$ expansions, the spatial dimension at which this level crossing
occurs can be estimated to be $d\approx3.4$, which means that the
three-fermion ground state at $d=3$ has $l=1$.  The same level crossing
has to occur for the five-fermion case about $d\approx3.64$, which
implies the five-fermion ground state with $l=1$ at $d=3$.  On the other
hand, the ground state of four or six fermions in a harmonic potential
has zero orbital angular momentum near $d=2$ and $d=4$.  Thus the level
crossing with higher orbital angular momentum states is unlikely and we
expect $l=0$ for the ground state at $d=3$.

In order to make quantitative discussions, we can use the Pad\'e
approximants to interpolate the two expansions around $d=2$ and $d=4$.
For each operator, we approximate its scaling dimension as a function of
$d=2+\bar\eps$ by a ratio of two polynomials;
\begin{equation}\label{Pade}
 [X/Y]=\frac{a_0+a_1\bar\eps+\cdots+a_X\bar\eps^{X}}
  {1+b_1\bar\eps+\cdots+b_Y\bar\eps^{Y}}.
\end{equation}
We demand that the series expansions of~(\ref{Pade}) around $d=2$ and
$d=4$ match the computed results.  $X+Y$ is fixed by the number of known
terms in the two expansions, while there is a freedom in distributing
the sum between $X$ and $Y$.

\begin{figure}[tp]
 \includegraphics[width=0.5\textwidth,clip]{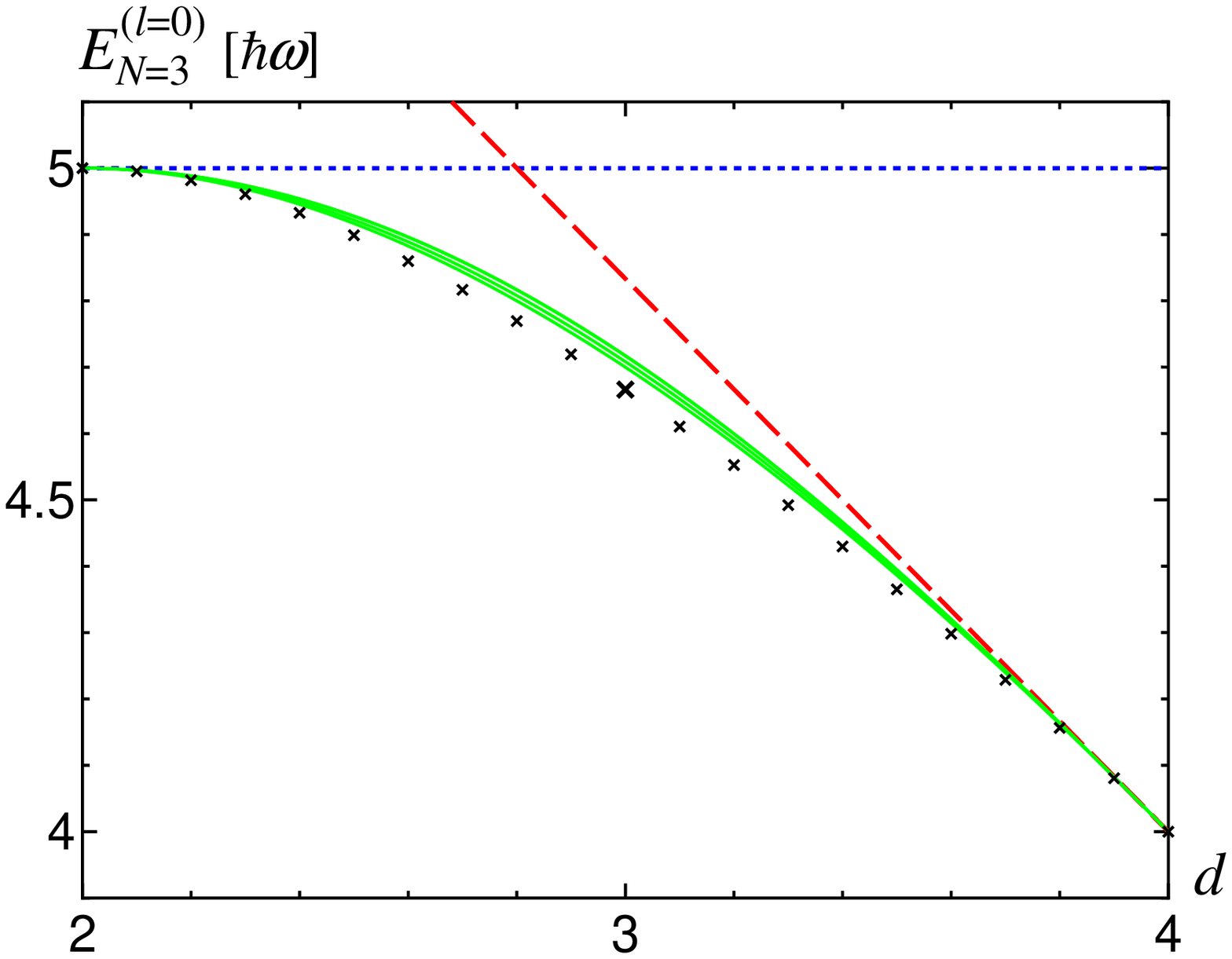}\hfill
 \includegraphics[width=0.5\textwidth,clip]{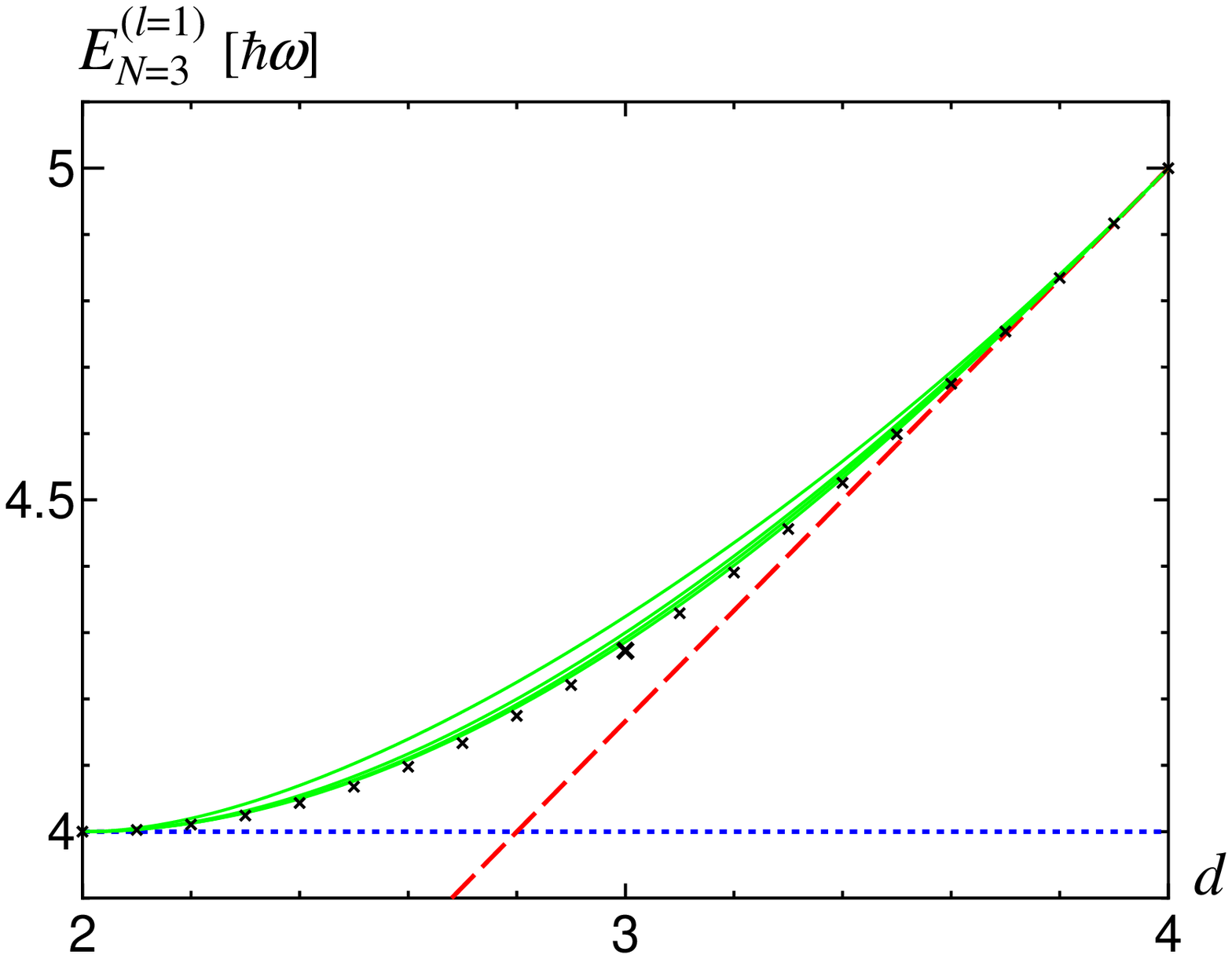}
 \caption{Energy of three fermions in a harmonic potential with $l=0$
 (left panel) and $l=1$ (right panel) as functions of spatial dimension
 $d$.  The four solid curves, although hard to distinguish, represent
 different Pad\'e approximants interpolating the expansions around $d=2$
 and $d=4$.  The dashed (dotted) lines are extrapolations from the
 $\eps=4-d$ ($\bar\eps=d-2$) expansions.  The symbols ($\times$)
 indicate the exact values for each $d$~\cite{Tan}.  \label{fig:Pade}}
\end{figure}

The different four Pad\'e approximants for the scaling dimension of each
three-fermion operator are plotted as functions of $d$ in
Fig.~\ref{fig:Pade}.  We find the behaviors of the Pad\'e approximants
are quite consistent with the exact results both for $l=0$ (left panel)
and $l=1$ (right panel).  For three-fermions with $l=0$, the
interpolated results at $d=3$ are
\begin{equation}
 \begin{split}
  [3/0]&=4.71,\qquad [2/1]=4.7,\\
  [1/2]&=4.72,\qquad [0/3]=4.72.
 \end{split}
\end{equation}
We see that all Pad\'e approximants give very close results in a small
interval $E_3^{(0)}\approx4.71\pm0.01$.  The harmonic oscillator
frequency $\omega$ was set to $1$ again.  This is close to the exact
result $4.66622$ at $d=3$~\cite{Tan04}, while the numbers obtained by
the Pad\'e interpolations are slight overestimates of the exact value.

Similarly, for three fermions with $l=1$, the interpolated results are
given by
\begin{equation}
 \begin{split}
  [3/0]&=4.29,\qquad [2/1]=4.3,\\
  [1/2]&=4.32,\qquad [0/3]=4.29,
 \end{split}
\end{equation}
which span a small interval $E_3^{(1)}\approx4.30\pm0.02$.  Again the
result is very close to, but slightly larger than the exact value
$4.27272$ at $d=3$~\cite{Tan04}.  One may expect that these agreements
will be further improved once additional terms in the expansions around
$d=2$ and $d=4$ are included.

For four fermions with $l=0$, the  Pad\'e interpolations to $d=3$ give
\begin{equation}
 \begin{split}
  [4/0]&=5.55,\qquad [3/1]=4.94,\qquad [2/2]=4.94,\\
  [1/3]&=4.90,\qquad [0/4]=6.17.
 \end{split}
\end{equation}
Some of these estimates are not too far from the numerical result
$5.1\pm0.1$~\cite{Chang07} and $5.07\pm0.01$~\cite{Stecher07} at $d=3$,
but the results span a larger interval $E_4^{(0)}\approx5.53\pm0.64$.
This may be because the coefficient of the next-to-leading-order
correction around $d=4$ is sizable compared to the leading term.  If one
excluded the two extremely asymmetric cases $[4/0]$ and $[0/4]$ where
all terms in the Pad\'e approximant come to the numerator or
denominator, one would have a rather small interval about
$E_4^{(0)}\approx4.92\pm0.02$.

For five fermions with $l=0$, the interpolated results at $d=3$ are
given by
\begin{equation}
 \begin{split}
  [3/0]&=7.71,\qquad [2/1]=7.64,\\
  [1/2]&=7.66,\qquad [0/3]=7.82,
 \end{split}
\end{equation}
which are in an interval $E_5^{(0)}\approx7.73\pm0.09$.  On the other
hand, for five fermions with $l=1$, the interpolated results are
\begin{equation}
 \begin{split}
  [3/0]&=7.10,\qquad [2/1]=7.16,\\
  [1/2]&=7.19,\qquad [0/3]=7.09,
 \end{split}
\end{equation}
which are in an interval $E_5^{(1)}\approx7.14\pm0.05$.  We find
$E_5^{(1)}<E_5^{(0)}$ at $d=3$, which means $l=1$ for the five-fermion
ground state in a harmonic potential.  However, its energy eigenvalue is
a substantial underestimate of the numerical result $7.6\pm0.1$ at
$d=3$~\cite{Chang07}.

For six fermions with $l=0$, the Pad\'e interpolations to $d=3$ give
\begin{equation}
 \begin{split}
  [4/0]&=10.1,\qquad [3/1]=7.92,\qquad [2/2]=7.92,\\
  [1/3]&=7.80,\qquad [0/4]=16.4.
 \end{split}
\end{equation}
The results now span a considerably larger interval
$E_6^{(0)}\approx12.1\pm4.3$, probably because of the huge
next-to-leading-order coefficient around $d=4$.  The large error signals
the worse convergence of the series expansions as the number of fermions
increases.  If the two extremely asymmetric cases $[4/0]$ and $[0/4]$
were excluded, one would have $E_6^{(0)}\approx7.86\pm0.06$.  For
comparison, the numerical result is $8.7\pm0.1$~\cite{Chang07} and
$8.67\pm0.03$~\cite{Stecher07} at $d=3$.

As the number of fermions goes to infinity, one can expect the series
expansions over $\bar\eps=d-2$ and $\eps=4-\eps$ for the ground state
energy break down.  Indeed, the energy of $N$ fermions at unitarity in
a harmonic potential scales with different powers of $N$ in different
spatial dimensions as $E_N\sim N^{(d+1)/d}$ for sufficiently large
$N$.  Therefore it is not surprising that our extrapolations to $d=3$
do not work well for five and six fermions.  It is possible that
the situation is improved once we know the next terms in the expansions
around $d=2$ and $d=4$.

Here we comment on the convergence of the $\bar\eps$ and $\eps$
expansions.  Since the exact integral equation to determine the energy
eigenvalues of three fermions in a harmonic potential is
known~\cite{Tan}, one can estimate the radii of convergence of the
expansions around $d=2$ and $d=4$ by studying their asymptotic
behaviors.  It turns out that the expansions for the three-fermion state
with $l=0$ are convergent when $|\bar\eps|\lesssim1.0$ or
$|\eps|\lesssim0.48$, while those with $l=1$ are convergent when
$|\bar\eps|\lesssim1.0$ or $|\eps|\lesssim1.4$.  On this basis, we
speculate that the expansions over $\bar\eps=d-2$ and $\eps=4-d$ have
nonzero radii of convergence for systems with a finite number of
particles.  The full details will be reported
elsewhere~\cite{Nishida-Tan}.

\section{Example 2: Anyons}

Anyons in two spatial dimensions present another example of a
nonrelativistic CFT.  In this section, we compute the scaling dimensions
of some operators near the bosonic limit and the fermionic limit, where
perturbative expansions in terms of statistical parameter $\theta$
are available.  Our analytical results, as we will see, are consistent
with results obtained by the conventional Rayleigh-Schr\"odinger
perturbation theory in a harmonic potential or numerical simulations.

The field-theoretical representation of anyons is provided by the
following Lagrangian density where a nonrelativistic field $\varphi$ is
minimally coupled to a Chern-Simons gauge field $a_\mu=(a_0,\a)$:
\begin{equation}
 \begin{split}
  \mathcal L &= 
  \frac1{4\theta}\partial_t\a\times\a-\frac1{2\theta}a_0\grad\times\a
  -\frac1{2\xi}\left(\grad\cdot\a\right)^2 \\
  &\qquad +i\varphi^*\left(\d_t+ia_0\right)\varphi
  -\frac1{2}\left|\left(\grad-i\a\right)\varphi\right|^2
  -\frac{v}4\left(\varphi^*\varphi\right)^2.
 \end{split}
\end{equation}
$\varphi$ is either a bosonic or fermionic field.  We denote the
propagator of $\varphi$ by $G(p)$. In the Coulomb gauge $\xi=0$, the
only nonvanishing components of the $a_\mu$ propagator are
\begin{equation}
  D_{i0}(p)=-D_{0i}(p)=-2i\theta\frac{\epsilon_{ij}p_j}{p^2}.
\end{equation}
We define the three-point vertex $\Gamma_0=-1$,
$\Gamma_i(p,p')=\left(p_i+p'_i\right)/2m$ and the four-point vertex
$\Gamma_{ij}=-\delta_{ij}/m$.  The contact interaction coupling $v$ has
to be fine-tuned so that the system is scale invariant.  We start with
the case where $\varphi$ is bosonic.

\subsection{Near the bosonic limit}

\subsubsection{Fixed points}

\begin{figure}[tp]
 \includegraphics[width=0.9\textwidth,clip]{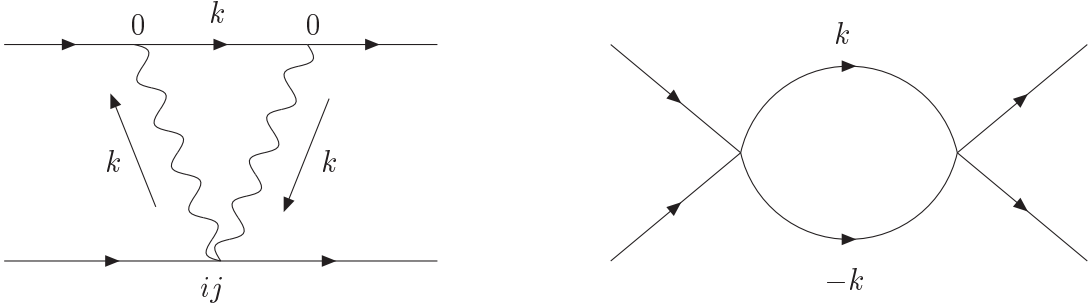}
 \caption{One-loop diagrams to renormalize the contact interaction
 coupling $v$.  \label{fig:anyon-vertex}}
\end{figure}

There are two one-loop diagrams which are logarithmically divergent and
renormalize the coupling $v$ (Fig.~\ref{fig:anyon-vertex}).
Integrating out modes in the momentum shell $e^{-s}\Lambda<k<\Lambda$,
the first diagram is evaluated as
\begin{equation}
 -4\int\!\frac{dk}{(2\pi)^3}
  G(k) \Gamma_0 D_{0i}(k) \Gamma_{ij} D_{j0}(k) \Gamma_0
  = -i\frac{4\,\theta^2}\pi \ln\frac\Lambda{e^{-s}\Lambda},
\end{equation}
while the second one as
\begin{equation}
 \begin{split}
  \frac{v^2}2 \int\!\frac{dk}{(2\pi)^3} G(k)G(-k)
  = i\frac{v^2}{4\pi}\ln\frac\Lambda{e^{-s}\Lambda}.
 \end{split}
\end{equation}
Therefore the renormalization group equation for $v$ is
\begin{equation}
 \frac{\d v}{\d s} = \frac{4\,\theta^2}\pi - \frac{v^2}{4\pi}.
\end{equation}
We find two fixed points located at~\cite{Bergman:1993kq}
\begin{equation}
 v=\pm4|\theta|.
\end{equation}
At these fixed points, the theory is a nonrelativistic CFT.  The
repulsive (upper sign) or attractive (lower sign) contact interaction
corresponds to a different boundary condition imposed on the $s$-wave
two-body wave function at origin $\sim
r^{\pm|\theta|/\pi}$~\cite{Manuel:1990in,Amelino-Camelia95}.

\subsubsection{Scaling dimensions of operators}

Since the one-body operator $\varphi$ is not renormalized, its scaling
dimension is $\Delta_\varphi = 1$, independent of $\theta$.  Using the
operator-state correspondence, we find that there is a one-anyon state
in a harmonic potential with energy $1\,\omega$, which is obvious.

\begin{figure}[tp]
 \includegraphics[width=0.8\textwidth,clip]{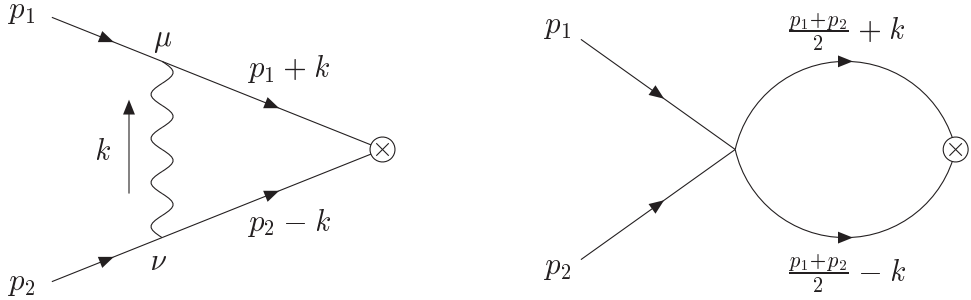}
 \caption{One-loop diagrams to renormalize the two-anyon operators.
 \label{fig:two-anyons}}
\end{figure}

The two-body operator $\varphi^2$ is renormalized by the diagrams
depicted in Fig.~\ref{fig:two-anyons}, which are potentially
logarithmically divergent and contribute to the anomalous dimension to
order $\theta$.  Since the first diagram turns out to be finite, we
can concentrate on the second diagram which is given by
\begin{equation}
 i\frac{v}2\int\!\frac{dk}{(2\pi)^3}G(k)G(-k)
  =-\frac{v}{4\pi}\ln\frac\Lambda{e^{-s}\Lambda}.
\end{equation}
Therefore the renormalized operator differs from the bare operator by
a rescaling factor: 
$(\varphi^2)_\mathrm{ren}=Z_{\varphi^2}^{-1}\varphi^2$, where
\begin{equation}
  Z_{\varphi^2} = 1 -\frac{v}{4\pi} s.
\end{equation}
At the fixed point, 
$\gamma_{\varphi^2}=-\d\ln Z_{\varphi^2}/\d s=\pm|\theta|/\pi$ is the
anomalous dimension of the composite operator $\varphi^2$.  We thus find
the scaling dimension
\begin{equation}
  \Delta_{\varphi^2} = 2\Delta_\varphi + \gamma_{\varphi^2}
  = 2 \pm \frac{|\theta|}\pi.
\end{equation}
According to the operator-state correspondence, $\varphi^2$ corresponds
to a two-anyon state in a harmonic potential with energy equal to
\begin{equation}
 E_2=\left[2 \pm \frac{|\theta|}\pi\right]\omega.
\end{equation}

It is straightforward to generalize our analysis to the lowest $N$-anyon
operator $\varphi^N$.  Its scaling dimension is given by
\begin{equation}
 \Delta_{\varphi^N} = N \pm \frac{N(N-1)}2 \frac{|\theta|}\pi,
\end{equation}
and therefore, the corresponding $N$-anyon state in a harmonic potential
has the energy $E_N=\Delta_{\varphi^N}\,\omega$.  This result coincides
with the exact energy eigenvalues of $N$ anyons in a harmonic
potential~\cite{Chou91,Basu92,Murthy92}.

\subsection{Near the fermionic limit}

\subsubsection{Scaling dimensions of operators}

If $\varphi$ is a fermionic field, the contact interaction term
$\left(\varphi^*\varphi\right)^2$ vanishes and the left diagram in
Fig.~\ref{fig:anyon-vertex} turns out to be finite.  Therefore, the
system is automatically scale invariant.  We denote the statistical
parameter in this case by $\theta'=\theta-\pi$.

Let us first look at the lowest two-body operators
$\varphi\nabla_{\!i\,}\varphi$.  Inserting this operator into the left
diagram of Fig.~\ref{fig:two-anyons}, its renormalization is given by
\begin{equation}
 \begin{split}
  &i\int\!\frac{dk}{(2\pi)^3}\left(\p_2-\k\right)_i G(p_1+k)
  \Gamma_\mu(p_1,p_1+k) D_{\mu\nu}(k) \Gamma_\nu(p_2,p_2-k) G(p_2-k)\\
  &-i\int\!\frac{dk}{(2\pi)^3}\left(\p_1+\k\right)_i G(p_1+k)
  \Gamma_\mu(p_1,p_1+k) D_{\mu\nu}(k) \Gamma_\nu(p_2,p_2-k) G(p_2-k)\\
  &=i\frac{\theta'}{\pi}\epsilon_{ij}\left(\p_2-\p_1\right)_j
  \ln\frac\Lambda{e^{-s}\Lambda}.
 \end{split}
\end{equation}
We thus find the linear combinations 
$\varphi\nabla_x\varphi\mp i\varphi\nabla_y\varphi$ with well-defined anomalous
dimensions $\gamma=\pm\theta'/\pi$.  Therefore, the scaling dimensions of
such operators are
\begin{equation}
 \Delta_{\varphi\nabla_x\varphi\mp i\varphi\nabla_y\varphi}
  = 3 \pm\frac{\theta'}\pi.
\end{equation}
The operator-state correspondence tells us that the two-anyon states in a
harmonic potential have the energies
\begin{equation}
 E_2 = \left[3 \pm\frac{\theta'}\pi\right]\omega.
\end{equation}
This result coincides with the exact energy eigenvalues of two anyons in
a harmonic potential.

Similarly, we can find the lowest three-body operator
$\varphi\nabla_x\varphi\nabla_y\varphi$ has a vanishing anomalous
dimension to order $\theta'$.  Therefore, the ground state energy of
three anyons in a harmonic potential near the fermionic limit is given by
\begin{equation}
 E_3 = \left[5+O(\theta'^2)\right]\omega.
\end{equation}
The same result has been derived using the conventional
Rayleigh-Schr\"odinger perturbation theory up to order
$\theta'^2$~\cite{Chou91}.

We now turn to the four-anyon case.  There are four lowest operators
that can mix with each other:
$\varphi\nabla_x\varphi\nabla_y\varphi\nabla_{xx}\varphi$,
$\varphi\nabla_x\varphi\nabla_y\varphi\nabla_{xy}\varphi$, 
$\varphi\nabla_x\varphi\nabla_y\varphi\nabla_{yy}\varphi$, and
$\varphi\nabla_x\varphi\nabla_y\varphi\d_t\varphi$.
The linear combinations with well-defined anomalous dimensions are
\begin{itemize}
\item $\varphi\nabla_x\varphi\nabla_y\varphi\d_t\varphi$ 
      with $\gamma=0$,
\item $\varphi\nabla_x\varphi\nabla_y\varphi\nabla^2\varphi$ 
      with $\gamma=0$,
\item $\varphi\nabla_x\varphi\nabla_y\varphi\left(\nabla_x-i\nabla_y\right)^2\varphi$   
      with $\gamma=\frac52\frac{\theta'}\pi$, 
\item $\varphi\nabla_x\varphi\nabla_y\varphi\left(\nabla_x+i\nabla_y\right)^2\varphi$
      with $\gamma=-\frac52\frac{\theta'}\pi$.
\end{itemize}
The combination of the first two operators
$\varphi\nabla_x\varphi\nabla_y\varphi\left(i\d_t+\tfrac12{\nabla^2}\right)\varphi$
annihilates the vacuum and thus does not correspond to any eigenstate of
the system in a harmonic potential.  The other three operators,
therefore, correspond to the energy eigenstates of four anyons in a
harmonic potential.  According to the operator-state correspondence, the
energies of these three states are given by
\begin{equation}
  E_4 = \left[8 + O(\theta'^2)\right]\omega \qquad\text{and}\qquad
  E_4 = \left[8 \pm\frac52\frac{\theta'}\pi
  + O(\theta'^2)\right]\omega.
\end{equation}
To our knowledge, these analytical results have not been derived so far.
Our numbers are consistent with slopes observed in the numerical
simulation~\cite{Sporre92}.

\section{Conclusion}

In this paper we study Schr\"odinger algebra and its representation in
terms of operators.  We show that irreducible representations are
built upon primary operators.  We also point out a correspondence
between primary operators and eigenstates in a harmonic potential.  We
illustrate this connection by computing the energy eigenvalues of up to
six fermions at unitarity in a harmonic potential using expansions over
$4-d$ and $d-2$, as well as the energy eigenvalues
of up to four anyons in a harmonic potential using expansions over
$\theta$ and $\theta-\pi$.

\acknowledgments

The authors thank Shina Tan for discussions.  Y.\,N.\ is supported by
JSPS Postdoctoral Fellowships for Research Abroad.
This work is supported, in part, by DOE Grant No.\ DE-FG02-00ER41132.

\appendix

\section{Commutators in the Schr\"odinger algebra}
\label{sec:algebra}

The Schr\"odinger algebra is formed from the operators $N$, $D$,
$M_{ij}$, $K_i$, $P_i$, $C$, $H$.  The commutators of $N$ and $M_{ij}$ with
other operators are
\begin{equation}
  [N,\, D] = [N,\, M_{ij}] = [N,\, K_i] = [N,\, P_i] 
  = [N,\, C]  = [N,\,H] = 0, 
\end{equation}
\begin{align}
  & [M_{ij},\, M_{kl}]= i(\delta_{ik} M_{jk} - \delta_{jk} M_{il}
    + \delta_{il} M_{kj} - \delta_{jl} M_{ki}),\\
  & [M_{ij},\, K_k] = i(\delta_{ik} K_j - \delta_{jk} K_i),\qquad
    [M_{ij},\, P_k] = i(\delta_{ik} P_j - \delta_{jk} P_i),\\
  & [M_{ij},\, C]  = [M_{ij},\, D] = [M_{ij},\, H] = 0.
\end{align}
The rest of the algebra is summarized in Table~\ref{tab:algebra}.

\begin{table}[tp]
\caption{Part of the Schr\"odinger algebra. Given are the values of
$[A,\,B]$. \label{tab:algebra}}
\begin{tabular*}{0.6\textwidth}%
   {@{\extracolsep{\fill}}|c|c|c|c|c|c|}
\hline
  \backslashbox{$A$}{$B$} & $P_j$ & $K_j$ & $D$ & $C$ & $H$ \\
\hline
  $P_i$ & $0$ & $-i\delta_{ij}N$ & $-iP_i$ & $-iK_i$ & $0$ \\
\hline
  $K_i$ & $i\delta_{ij}N$ & $0$ & $iK_i$ & $0$ & $iP_i$ \\
\hline
  $D$ & $iP_j$ & $-iK_j$ & $0$ & $-2iC$ & $2iH$ \\
\hline
  $C$ & $iK_j$ & $0$ & $2iC$ & $0$ & $iD$ \\
\hline
  $H$ & $0$ & $-iP_j$ & $-2iH$ & $-iD$ & $0$ \\
\hline
\end{tabular*}
\end{table}

\section{The stress tensor}
\label{sec:stress}

We will find the stress tensor $\Pi_{ij}$ which appears in the
momentum conservation equation:
\begin{equation}\label{mom-cons}
  \d_t j_i + \d_j \Pi_{ij} = 0.
\end{equation}
From the evolution equation for the field operators $\psi$ and
$\psi^\+$ it follows that
\begin{multline}\label{dmj}
  \d_t (\j(\x)) = \frac1{4} (\psi^\+\nabla^2\grad\psi 
  + \nabla^2\grad\psi^\+\psi - \nabla^2\psi^\+\grad\psi
  - \grad\psi^\+\nabla^2\psi)(\x) \\
    - \int\!d\y\, \grad_{\x}V(\x-\y) :\!n(\x)n(\y)\!:,
\end{multline}
where $:\!\cdots:\!$ denotes normal ordering [i.e.,
$:\!n(\x)n(\y)\!:
=\psi^\+_\alpha(\x)\psi^\+_\beta(\y)\psi_\beta(\y)\psi_\alpha(\x)$].
It is not obvious that the right hand side of Eq.~(\ref{dmj}) can be
written as the derivative of a stress tensor.  To do that, let us
introduce the Laplace transform $\alpha(Q)$ of the function $rV(r)$:
\begin{equation}
  4\pi r V(r) = \int\limits_0^\infty\!dQ\, \alpha(Q) e^{-Qr}.
\end{equation}
In other words, we write the potential $V(r)$ as a superposition of
Yukawa potentials,
\begin{equation}
  V(r) = \int\limits_0^\infty\!dQ\, \alpha(Q) \frac{e^{-Qr}}{4\pi r}\,.
\end{equation}
We also introduce, for each value of $Q$, an auxiliary field
$\sigma_Q(\x)$,
\begin{equation}
  \sigma_Q(\x) = \int\!d\y\, \frac{e^{-Q|\x-\y|}}{4\pi|\x-\y|}n(\y).
\end{equation}
It satisfies the equation
\begin{equation}\label{sigma-eq}
  (-\nabla^2+Q^2) \sigma_Q(\x) = n(\x).
\end{equation}
The stress tensor can now be introduced: 
\begin{multline}
  \Pi_{ij} = \frac1{2} (\d_i\psi^\+\d_j\psi + \d_j\psi^\+\d_i\psi)
    - \frac1{4}\delta_{ij}\nabla^2 n\\
    + \int\limits_0^\infty\!dQ\, \alpha(Q) :\!\left\{
  -\d_i\sigma_Q\d_j\sigma_Q + \frac{\delta_{ij}}2
  \left[(\nabla\sigma_Q)^2 + Q^2\sigma_Q^2\right]\right\}\!:.
\end{multline}
By using Eq.~(\ref{sigma-eq}) it is straightforward to verify that
Eq.~(\ref{mom-cons}) is satisfied.

Notice that $\Pi_{ij}$ is not unique.  For example, one can replace
\begin{equation}
  \Pi_{ij} \to \Pi_{ij} + (\d_i\d_j - \delta_{ij}\nabla^2) \Phi
\end{equation}
with any $\Phi$ without destroying the momentum conservation.

Let us now show that if $V(r)$ is a short-range potential with infinite
scattering length, then
\begin{equation}\label{intP}
  \int\!d\x\, \Pi_{ii}(\x) = 2H.
\end{equation}
By using Eq.~(\ref{sigma-eq}) and the following property of the Yukawa
potential,
\begin{equation}
  \int\!d\x\, \frac{e^{-Q|\x-\y|}}{4\pi|\x-\y|}\, 
  \frac{e^{-Q|\x-\z|}}{4\pi|\x-\z|} 
  = \frac{e^{-Q|\y-\z|}}{8\pi Q} 
\end{equation}
(which can be shown, e.g., by using the Fourier transforms), we find
\begin{equation}
  \int\!d\x\, \Pi_{ii}(\x) = 2T + V + \int\!dQ\,d\x\,d\y\,
  \alpha(Q) Q \frac{e^{-Q|\x-\y|}}{8\pi}:\!n(\x)n(\y)\!:,
\end{equation}
where $T$ is the kinetic energy and $V$ is the potential energy.
Since $V(r)$ is a short-range potential giving an infinite scattering
length, the low-energy physics does not change when one rescales the
potential as 
\begin{equation}\label{Vtr}
 V(r) \to \lambda^2 V(\lambda r).
\end{equation}
In particular the Hamiltonian is unchanged under the
transformation~(\ref{Vtr}).  Setting $\lambda=1+\epsilon$,
$\epsilon\ll1$ and expanding $H$ to the linear order in $\epsilon$, we find
\begin{equation}
  \int\!dQ\,d\x\,d\y\, \alpha(Q) Q 
  \frac{e^{-Q|\x-\y|}}{8\pi}:\!n(\x)n(\y)\!: \,= V.
\end{equation}
Therefore, we obtain Eq.~(\ref{intP}).

One can use this relationship to prove that in the normal phase (above
the critical temperature), the bulk viscosity of a Fermi gas at
unitarity is identically zero.  Indeed, the bulk viscosity is given by
the Kubo's formula:
\begin{equation}
  \zeta = \lim_{\omega\to0}\frac1{9\omega}
     \int\limits_0^\infty\!dt\!\int\!d\x\, e^{i\omega t} 
          \<[\Pi_{ii}(t,\x),\, \Pi_{jj}(0,\bm{0})]\>,
\end{equation}
but the integral over $\x$ can be taken according to Eq.~(\ref{intP}).
Moreover, as $\<[H,\, \O]\>=0$ in thermal equilibrium for any operator
$\O$, the bulk viscosity is zero.  This result was derived previously
using a different approach~\cite{Son:2005tj}.




\begin{thebibliography}{99}

\bibitem{Fradkin:1997df}
  See, e.g., E.~S.~Fradkin and M.~Y.~Palchik,
  ``New developments in $D$-dimensional conformal quantum field theory,''
  Phys.\ Rept.\ {\bf 300}, 1 (1998).

\bibitem{Mehen:1999nd}
  T.~Mehen, I.~W.~Stewart, and M.~B.~Wise,
  ``Conformal invariance for non-relativistic field theory,''
  Phys.\ Lett.\  B {\bf 474}, 145 (2000)
  [arXiv:hep-th/9910025].

\bibitem{Hagen:1972pd}
  C.~R.~Hagen,
  ``Scale and conformal transformations in Galilean-covariant field theory,''
  Phys.\ Rev.\  D {\bf 5}, 377 (1972).

\bibitem{Niederer:1972}
  U.~Niederer, 
  ``The maximal kinematical invariance group of the free Schr\"odinger 
   equation,'' 
  Helv.\ Phys.\ Acta {\bf 45}, 802 (1972).

\bibitem{Hussin:1986cc}
  V.~Hussin and M.~Jacques,
  ``On nonrelativistic conformal symmetries and invariant tensor fields,''
  J.\ Phys.\ A  {\bf 19}, 3471 (1986).

\bibitem{Jackiw:1990mb}
  R.~Jackiw and S.~Y.~Pi,
  ``Classical and quantal nonrelativistic Chern-Simons theory,''
  Phys.\ Rev.\  D {\bf 42}, 3500 (1990)
  [Erratum: \textit{ibid.}\ {\bf 48}, 3929 (1993)].

  R.~Jackiw and S.~Y.~Pi,
  ``Finite and infinite symmetries in (2+1)-dimensional field theory,''
  arXiv:hep-th/9206092.

\bibitem{Henkel:1993sg}
  M.~Henkel,
  ``Schr\"odinger invariance in strongly anisotropic critical systems,''
  J.\ Statist.\ Phys.\  {\bf 75}, 1023 (1994)
  [arXiv:hep-th/9310081].

\bibitem{WernerCastin}
  F.~Werner and Y.~Castin,
  ``Unitary gas in an isotropic harmonic trap: symmetry properties and
  applications,''
  Phys.\ Rev.\ A {\bf 74}, 053604 (2006)
 [arXiv:cond-mat/0607821].

\bibitem{Tan}
 S.~Tan, private communication. 

\bibitem{Nishida-Son06}
  Y.~Nishida and D.~T.~Son,
  ``$\epsilon$ expansion for a Fermi gas at infinite scattering length,''
  Phys.\ Rev.\ Lett.\ {\bf 97}, 050403 (2006)
  [arXiv:cond-mat/0604500].

  Y.~Nishida and D.~T.~Son, 
  ``Fermi gas near unitarity around four and two spatial dimensions,''
  Phys.\ Rev.\ A {\bf 75}, 063617 (2007)
  [arXiv:cond-mat/0607835].

\bibitem{Nikolic-Sachdev07}
  P.~Nikoli\'c and S.~Sachdev,
  ``Renormalization-group fixed points, universal phase diagram, and 
  $1/N$ expansion for quantum liquids with interactions near the 
  unitarity limit,''
  Phys.\ Rev.\ A {\bf 75}, 033608 (2007)
  [arXiv:cond-mat/0609106].

\bibitem{Nussinov06}
 Z.~Nussinov and S.~Nussinov, 
 ``Triviality of the BCS-BEC crossover in extended dimensions:
 Implications for the ground state energy,'' 
 \PRA{74,053622,2006}
 [arXiv:cond-mat/0410597].

\bibitem{Tan04}
  S.~Tan, 
  ``Short range scaling laws of quantum gases with contact interactions,''
  arXiv:cond-mat/0412764.

\bibitem{Chang07}
  S.~Y.~Chang and G.~F.~Bertsch, 
  ``Unitary Fermi gas in a harmonic trap,''
  \PRA{76,021603(R),2007}
  [arXiv:physics/0703190].

\bibitem{Stecher07}
  J.~von~Stecher, C.~H.~Greene, and D.~Blume,
  ``BEC-BCS crossover of a trapped two-component Fermi gas with unequal masses,''
  arXiv:0705.0671 [cond-mat.other].

\bibitem{Nishida-Tan}
  Y.~Nishida and S.~Tan, unpublished.

\bibitem{Bergman:1993kq}
  O.~Bergman and G.~Lozano,
  ``Aharonov-Bohm scattering, contact interactions, and scale invariance,''
  Annals Phys.\  {\bf 229}, 416 (1994)
  [arXiv:hep-th/9302116].

\bibitem{Manuel:1990in}
  C.~Manuel and R.~Tarrach,
  ``Contact interactions of anyons,''
  Phys.\ Lett.\  B {\bf 268}, 222 (1991).

\bibitem{Amelino-Camelia95}
 G.~Amelino-Camelia and D.~Bak, 
 ``Schr\"odinger self-adjoint extension and quantum field theory,''
 \PLB{343,231,1995}
 [arXiv:hep-th/9406213]..

\bibitem{Chou91}
 C.~Chou, 
 ``Multianyon spectra and wave functions,''
 \PRD{44,2533,1991} [Erratum: \ibid{45,1433,1992}].

\bibitem{Basu92}
 R.~Basu, G.~Date, and M.~V.~N.~Murthy, 
 ``Class of exact solutions for many-anyon quantum mechanics,''
 \PRB{46,3139,1992}.
 
\bibitem{Murthy92}
 M.~V.~N.~Murthy, J.~Law, R.~K.~Bhaduri, and G.~Date,
 ``On a class of non-interpolating solutions of the many-anyon problem,''
 J.\ Phys.\ A \journal{25,6163,1992}.

\bibitem{Sporre92}
 M.~Sporre, J.~J.~M.~Verbaarschot, and I.~Zahed, 
 ``Four anyons in a harmonic well,''
 \PRB{46,5738,1992}.

\bibitem{Son:2005tj}
  D.~T.~Son,
  ``Vanishing bulk viscosities and conformal invariance of the unitary 
  Fermi gas,''
  Phys.\ Rev.\ Lett.\  {\bf 98}, 020604 (2007)
  [arXiv:cond-mat/0511721].

\end{thebibliography}
\end{document}